\documentclass[12pt]{elsarticle}
\usepackage{amssymb,amsmath, bm,mathtools,xcolor, float}
\numberwithin{equation}{section}
\usepackage[ruled,linesnumbered,vlined]{algorithm2e}
\usepackage{graphicx}
\usepackage[colorlinks=true,linkcolor=blue]{hyperref}
\usepackage{geometry}
\usepackage{subcaption}
\usepackage{multirow}
\begin{document}
\captionsetup[subfigure]{skip=1pt, singlelinecheck=false}

\title{Geometry-mediated shear softening in dense ordered granular packings}
%\title{Shearing dense and ordered granular packings in confined space can be easy: A geometry-induced elastic anisotropy and softening}
\author[1,2]{Liuchi Li\corref{cor1}}
\cortext[cor1]{Corresponding author}
\ead{ll2756@princeton.edu.edu}
\author[3]{Konstantinos Karapiperis}
\address[1]{Department of Civil and Environmental Engineering, Princeton, NJ 08544, USA}
\address[2]{Princeton Materials Institute, Princeton University, Princeton, NJ 08544, USA}
\address[3]{School of Architecture, Civil and Environmental
Engineering, EPFL, Lausanne, Switzerland}

\begin{abstract}
Shearing a packing of solid granular grains can be difficult, especially when the solid fraction is high and the boundary confinement is strong. It was recently shown that embedding voids in grains can make a packing easier to shear when such voids make the grains auxetic. Here, we use finite element simulation to show that auxeticity is not a necessary condition even in a seemingly very constrained setting: shearing dense and ordered granular packings under a constant solid fraction. More specifically, by controlling the geometry of a void embedded in a grain, we induce an apparent elastic anisotropy and softening of the grain under shear, which collectively leads to a significant reduction -- up to 90\% -- of the apparent shear modulus of a packing of these grains. Complementary analysis shows that this reduction correlates well with a decrease in contact-force anisotropy, and is insensitive to system size and contact friction variation. Our results highlight how grain-scale geometry, mediated by multi-body contact mechanics, modulates macroscopic system-scale elasticity, providing a minimal design mechanism towards targeted collective mechanical properties of soft granular metamaterials.
\end{abstract}

 \maketitle

\section{Introduction}
The macroscopic behavior of granular materials emerges from contact interactions among discrete grains. Even when the constituent grains are geometrically ``simple'' and deform negligibly, their assembly can display a rich collective response, including phase transition between solid-like and fluid-like states under shear. The direction of this transition depends strongly on the boundary conditions. Under weak confinement, as in open-channel flows~\cite{forterre2008flows}, shearing can promote fluidization. On the contrary, when confinement is strong, e.g., when sheared in a spatially-bounded cell~\cite{bi2011jamming}, shearing can lead to jamming.

Recent studies have shown that this picture can change when the grains themselves can undergo finite deformation~\cite{cantor2020compaction,cardenas2022experimental}, challenging our current understanding of phase transition. For example, densely packed soft grains can become jammed and difficult to shear~\cite{poincloux2024rigidity}, while they can conversely become much easier to shear and fluidize if these grains were further made auxetic~\cite{haver2024elasticity}. The physical interpretation becomes clear by looking at what happens at the grain scale: for the former, finite grain deformation expands inter-grain contacts, suppresses structural rearrangements, and collectively leads to the emergence of rigidity; for the latter, finite grain deformation, on the contrary, diminishes inter-grain contacts, promotes structural rearrangements, and collectively leads to enhanced flowability.

Here, we numerically show that finite deformation, and more specifically auxeticity, is not a necessary condition. Under weak confinement, as in open-channel flows~\cite{forterre2008flows}, shearing can promote fluidization. ordered granular packings as in \cite{poincloux2024rigidity} under a constant solid fraction. More specifically, starting from a two-dimensional annular grain with a circular internal void, we perturb \textit{only} the shape of the void while preserving its area. This minimal geometric change induces an apparent elastic anisotropy and softening of the grain. When many such grains interact through contact, this grain-scale softening leads to a substantial reduction in the apparent shear modulus of the packing. Our analysis indicates that this reduction is correlated with a decrease in the anisotropy of the contact-force network and remains robust against variations in system size and frictional properties. These findings suggest a minimal design route for contact-driven soft granular metamaterials: by tuning geometry at a smaller length scale, one can program mechanical response at a larger length scale.
 
 The rest of the paper is organized as follows: in Section~\ref{setup}, we discuss the design of grains with embedded anisotropic voids and the setup of our virtual experiments, in Section~\ref{approach}, we discuss briefly the numerical modeling approach, in Section~\ref{result}, we discuss the results, and lastly, in Section~\ref{discussion}, we present a summary and a discussion for future work.

\section {Model system and shear protocol} \label{setup}

\begin{figure}[h]
\centering
\includegraphics[width=\linewidth]{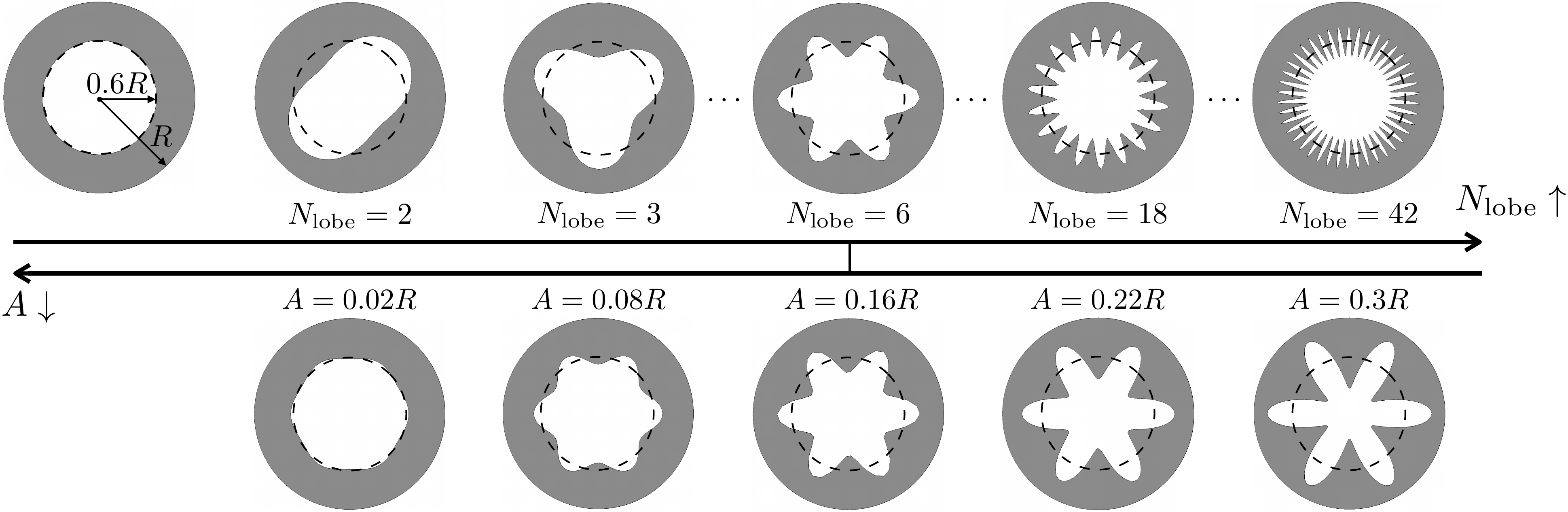}
\caption{A schematic showing the design of a void's geometry, parametrized by $N_\text{lobe}$ and $A$. Note that all void designs occupy the same area: $0.36\pi R^2$.}
\label{grain-design}
\end{figure}

We start from the 2D ``donut'' grain considered in \cite{poincloux2024rigidity} as our reference design. It is a circular grain of radius $R = 5$ mm containing an inner radius of $0.6R$. To isolate the effect of the void shape, we keep the void area constant and only perturb the void boundary. More specifically, this boundary is described by a sinusoidal curve in polar coordinates, controlled by an amplitude $A$ and a frequency defined by the number of lobes $N_\text{lobe}$. Figure~\ref{grain-design} shows some examples of the designs we consider in this work, including the reference design ($N_\text{lobe}$ = 0, $A$ = 0), and a ``peanut''-shaped design ($N_\text{lobe} = 2$, $A$ = 0.16R). As the void's shape deviates from being isotropic (e.g., becoming a ``peanut''), the apparent elasticity of the grain will likely no longer be isotropic, giving rise to a geometry-induced anisotropy as a function of the void's angular orientation, which we denote as $\theta$ herein. Intuitively, we would expect the elastic anisotropy of the grain to increase first and later decrease, converging to being isotropic, as $N_\text{lobe}\rightarrow \infty$. As to the choice of the constituent material, we follow our previous work~\cite{li2021emerging}, and consider a rubber-like one with a Young's modulus of 5 MPa, a Poisson's ratio of $\frac{1}{3}$ describing a ``plane strain"-equivalent incompressible material~\cite{vu2019numerical}, and a material density of 1180 kg/m$^3$.  
 
\begin{figure} [h]
\centering
\includegraphics[width=\linewidth]{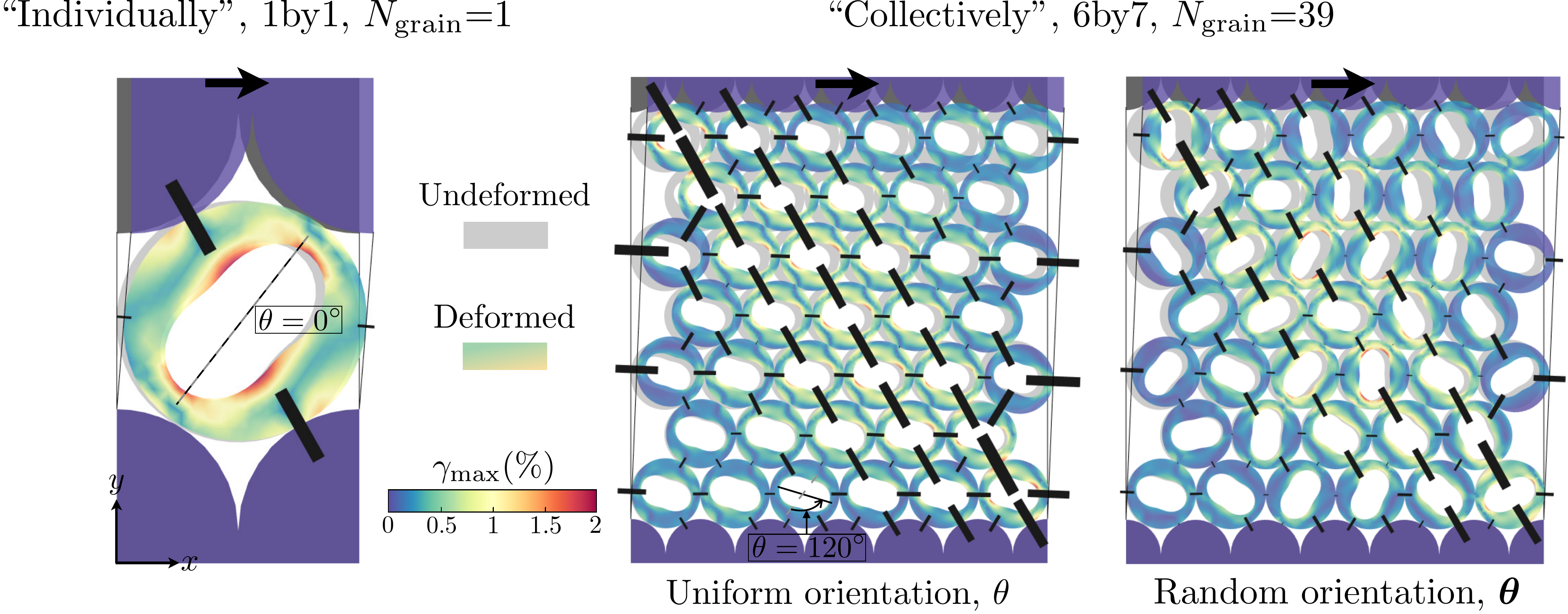}
\caption{Schematics for our virtual shear experiment. (a) The undeformed and deformed configuration of a ``peanut'' grain ($N_\text{lobe}=2$, $A = 0.16R$) oriented at $\theta = 0^\circ$. The color map shows the spatial distribution of the maximum in-plane strain within the grain. We displace the top platen with a constant horizontal speed $V_x$ = 0.1 m/s, during which we collect $F_x$ by summing up contact forces (shown as black lines) involving the top platen. (b) Two similar schematics to that shown in (a), but for a collection (6 columns, 7 rows, $N_\text{grain} = 39$) of peanut grains. The left panel shows a scenario where all grains are oriented identically with $\theta=120^\circ$, while the right panel shows a scenario where all grains have different orientations, each of which is sampled independently from the same uniform distribution between $0^\circ$ and $180^\circ$. Contact forces (black lines) in both panels are plotted with the same scale for a direct visual comparison.}
\label{shear-schematic}
\end{figure}

We proceed to study the behavior of a hexagonal lattice of $N_\text{grain}$ \textit{metagrains} under shear, as shown in Figure~\ref {shear-schematic}. Note that this corresponds to their densest possible packing in the undeformed configuration. The special case of $N_\text{grain} = 1$ corresponds to the shearing of a single grain. A small amount of vertical pressure (1 MPa) is first applied through rigid platens made up of disks of the same radius as the grains. Once static equilibrium is achieved, the top platen is displaced horizontally with a constant speed $V_x$ = 0.1 m/s, without changing the volume of the shear cell. Given also the incompressible constituent material of the grains, this represents a shear protocol under constant solid fraction. We focus on the small-strain regime and impose shear strains up to a shear strain $\gamma$=3\%.

The effective shear modulus of the system is computed from the horizontal force exerted on the top platen:
\begin{align*}
G = \frac{F_x}{L_x\gamma},
\end{align*}
where $F_x$ is obtained by summing the horizontal components of the contact forces on the top platen, $L_x$ is the width of the shear cell, and $\gamma$ is the shear strain. In all simulations, $F_x$ varies approximately linearly with $\gamma$ after a short initial relaxation. We therefore determine $G$ from the slope of a linear fit to the force–strain data for $\gamma \geq 0.5$. 

Because the void shape can make each grain elastically anisotropic, the angular orientation of the grains becomes an additional control parameter. We consider two orientation protocols. In the first, all grains in a packing have the same orientation $\theta$, which we vary from $0^\circ$ to $360^\circ$ in increments of $12^\circ$. In the second, each grain is assigned a random orientation. For grains with $N_\mathrm{lobe}>0$, each orientation is sampled uniformly from $0^\circ$ to $360^\circ/N_\mathrm{lobe}$, and five independent random configurations are generated for each design. As an example, Figure~\ref{shear-schematic}(b) shows a hexagonal packing of 39 ``peanut'' grains with either a uniform orientation ($\theta = 120^\circ$, left panel) or random orientations (right panel). It is worth noting that, due to the elastic anisotropy of each grain, the granular packing is geometrically ordered but mechanically disordered, which can similarly give rise to different contact force distributions (see e.g. Figure~\ref{shear-schematic}(b)).

To investigate the connection between the particle-scale geometry-induced anisotropy (i.e., $N_\text{lobe}$, $A$, and $\theta$) and the effective shear modulus of the packing, we carry out two campaigns. The first campaign fixes $A=0.16R$ and varies $N_\mathrm{lobe}, N_\mathrm{grain}$, and the uniform orientation $\theta$. This set of simulations is used to examine how void shape affects both single-grain anisotropy and collective packing response. The second campaign focuses on randomly oriented 39-grain packings and varies both $A$ and $N_\mathrm{lobe}$. This is used to assess how strongly void-shape amplitude controls the shear modulus and to connect the macroscopic response to the contact-force network.

Henceforth, to indicate the shear modulus of a system with parameters $A, N_{\text{lobe}}, N_{\text{grain}}$, we use the notation $G(A, N_{\text{lobe}}, N_{\text{grain}}, \theta)$ in the uniform orientation scenario, and $G(A, N_{\text{lobe}}, N_{\text{grain}}, \boldsymbol{\theta})$ in the random orientation scenario. We consider $N_\text{lobe}\in\{0, 2, 3, 4, 5, 6, 7, 8, 9, 10, 11, 12, 18, 26, 42\}$ and $N_\text{grain}\in\{1, 5, 39, 137\}$. Note that $N_\text{lobe} = 0$ and $A=0$ occur simultaneously. To compare different designs, we report a normalized apparent shear modulus $\tilde{G}=G/G_{\text{ref}}$
where $G_\mathrm{ref} = \langle G(A, N_{\text{lobe}}, N_{\text{grain}}, \boldsymbol{\theta}) \rangle_{\boldsymbol{\theta}}$ is the ensemble-averaged shear modulus of the corresponding reference system. For uniformly oriented systems, the ensemble average is taken over the 30 values of $\theta$. For randomly oriented systems, the ensemble average is taken over the five independent random orientation configurations. In the following, values of $\tilde{G}<1$ indicate that the designed grains are easier to shear than the circular-void reference grains.

\section{Numerical method}\label{approach}
We utilize a multi-body contact mechanics approach~\cite{simo1992augmented,laursen2003computational} that conducts simulations through a standard finite element computation, based on our previous work~\cite{li2021emerging}, which investigated the heterogeneous contact force distribution arising from geometrically ordered but mechanically disordered granular packing under (quasi-)static isotropic compression. In essence, we extend the approach documented in~\cite{li2021emerging} to elastodynamics via the average acceleration (i.e., an implicit time integration) scheme, incorporating a Rayleigh damping term to account for any dissipative effect that will inevitably occur in reality. We restrict our attention to linear elasticity, meaning that the stiffness matrix associated with each grain remains constant. Mathematically, after a finite element discretization, we numerically solve the following linear system (other with any applicable boundary conditions):
\begin{align}
\mathbf{M}\ddot{\mathbf{U}}+\mathbf{C}\dot{\mathbf{U}}+\mathbf{K}\mathbf{U} = \mathbf{F}_\text{ext}+\mathbf{F}_\text{contact}(\mathbf{U}),
\end{align}
with
\begin{align}
\mathbf{C} = \alpha\mathbf{M}+\beta\mathbf{K},
\end{align}
where $\mathbf{M}$, $\mathbf{C}$, $\mathbf{K}$ are the block-diagonal, consistent mass matrix, damping matrix, and stiffness matrix, respectively,  considering all grains in a system. $\mathbf{U}$, $F_\text{ext}$, and $\mathbf{F}_\text{contact}$ are, respectively, the displacement vector, external force vector, and contact force vector, again, considering all grains in a system. $\mathbf{F}_\text{ext}$ accounts for the applied pressure term in the compression stage and becomes zero in the shear stage, while $\mathbf{F}_\text{contact}(\mathbf{U})$ accounts for the contact tractions among grains and depends on the solution, $\mathbf{U}$. $\alpha$ and $\beta$ are user-defined constants, and we set $\alpha = 0$ s$^{-1}$ and $\beta = 1\times10^{-3}$ s\footnote{We verified that results are not sensitive to different values of $\alpha$ (between 0 s$^{-1}$ and 30 s$^{-1}$) and $\beta$ (between $1\times10^{-2}$ s and $1\times10^{-4}$ s).} for all simulations utilizing a time step of $4\times10^{-5}$ s, which, together with the shear rate $V_x$ = 0.1 m/s, represents an overdamped dynamics and ensures an overall quasi-static response for the setup considered in this work. At each time step, we employ a Newton-Raphson solver to find, through iterations, a converged solution of $\mathbf{U}$ before moving to the next time step. Once we find $\mathbf{U}$, we can compute all nodal forces and subsequently contact forces by considering FEM nodes that are in contact. All other modeling-related parameters, such as the contact penalty constants, are the same as in~\cite{li2021emerging}, and we also treat the top and bottom platens, as well as the two vertical walls, as rigid objects. 

Finally, we consider Coulomb friction, parametrized by a constant $\mu_\text{p}$, to account for any friction that occurs in both grain-grain and grain-boundary interactions. In this work, we consider mainly frictionless interactions. In practice, we select $\mu_\text{p}= 0.001$, small yet not zero, to represent friction that may inevitably occur in reality.

\section{Results and discussion}\label{result}
\subsection{Uniformly oriented grains: Softening, anisotropy, and collectivity}\label{result_uniform}
\begin{figure}[h]
\centering
\includegraphics[width=\linewidth]{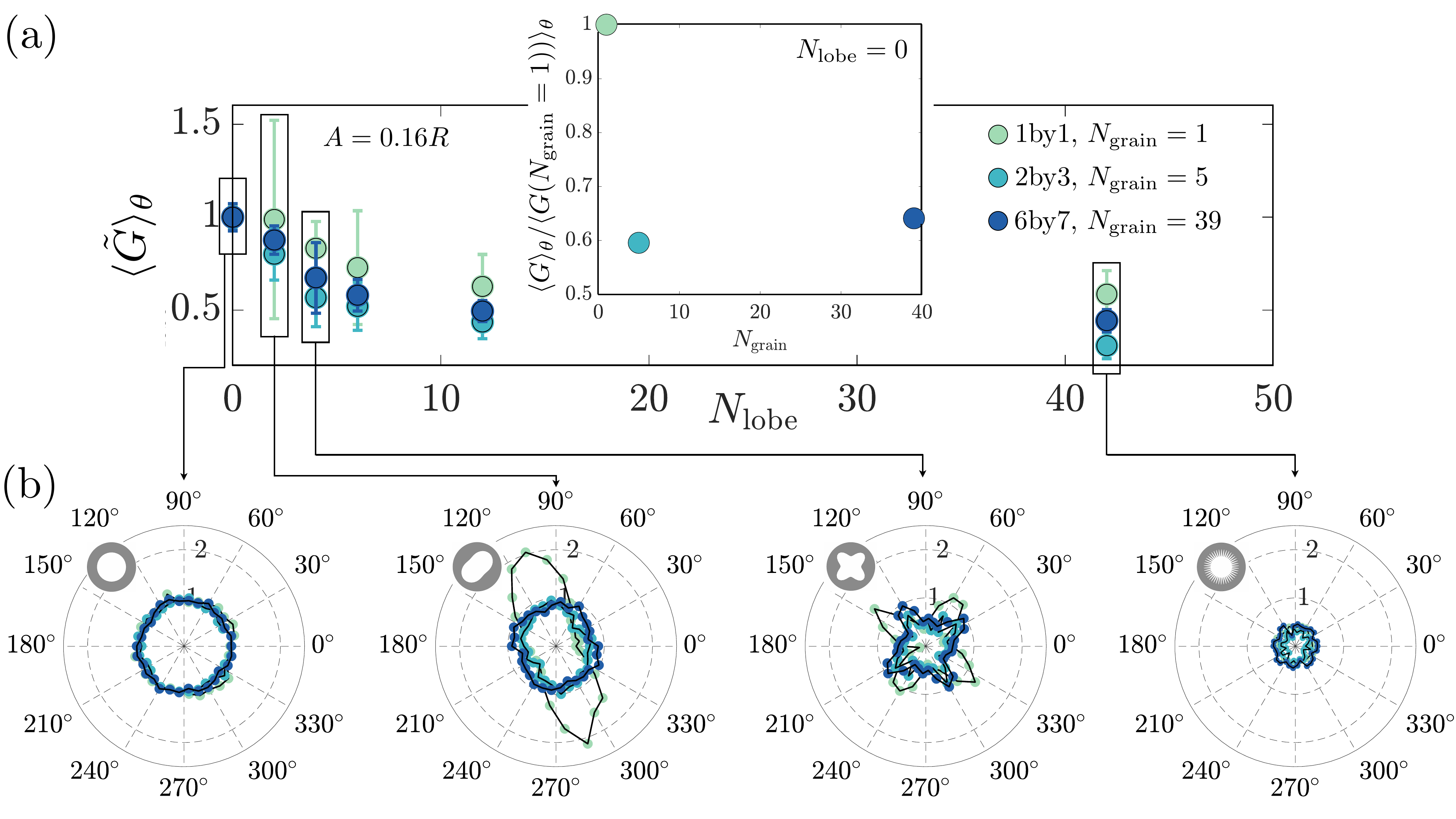}
\caption{(a) The variation of $\langle\tilde{G}(\cdot, N_\text{lobe}, N_\text{grain}, \theta)\rangle_\theta$, obtained by averaging over 30 configurations parametrized by $\theta$, as a function of $N_\text{lobe}$ under a fixed $A = 0.16R$. Error bars suggest the degree of anisotropy of $\tilde{G}(\cdot, N_\text{lobe}, N_\text{grain}, \theta)$ as a function of $\theta$. Results are shown for three system sizes : $N_\text{grain}=1$ (light green), $N_\text{grain}=5$ (cyan), and $N_\text{grain}=39$ (blue). The inset shows the value of $\langle G\rangle_\theta$ for the reference design ($N_\text{lobe} = 0$) as a function of $N_\text{grain}$, normalized by the case with $N_\text{grain} = 1$. (b) The angular distribution of $\tilde{G}(\cdot, N_\text{lobe}, N_\text{grain}, \theta)$ as a function of $\theta$, considering the same three system sizes, for $N_\text{lobe}$ = 0, 2, 4, and 42, respectively, from left to right.}
\label{anisotropy-uniform}
\end{figure}

We first focus on packings with uniform void orientation, and examine the role of lode number $N_{\text{lobe}}$ by fixing the amplitude $A$. Figure~\ref{anisotropy-uniform}(a) shows for different systems sizes ($N_\text{grain} = 1,5,39$) the variation of $\langle\tilde{G}\rangle_\theta$ as a function of $N_\text{lobe}$. Here $\langle \cdot \rangle$ indicates the ensemble average over 30 different angles $\theta$. The results show an elastic softening effect since $\langle\tilde{G}\rangle_\theta$ quickly decreases and converges to a value of around 0.5 as $N_\text{lobe}$ goes beyond 10. This observation remains in place regardless of the system size, for all three sizes considered. 

The second observation is that the error bars associated with data from $N_\text{lobe} < 10$ are considerable, especially more so for the single-grain system than for the other two multi-grain systems, indicating a considerable level of elastic anisotropy of a single grain with $N_\text{lobe}<10$ when sheared individually, which, however, becomes less pronounced when sheared collectively. Notably, with $N_\text{lobe}=2$, one ``peanut'' grain can be either much harder or much easier to shear than one ``donut'' grain, depending on its orientation $\theta$; however, a collection of ``peanut'' grains is always easier to shear than a collection of ``donut'' grains, regardless of the value of $\theta$. In other words, collectivity does not simply preserve the single-grain anisotropy; instead, multi-body contact interactions partially average it out.

The third observation is that collectivity changes the magnitude of the elastic softening. Once more than one grain is present, the packing response is softer than the corresponding single-grain response. This can already be seen in the circular-void reference systems shown in the inset of Figure 3(a), where the moduli of the 5-grain and 39-grain systems are both about 60\% of the single-grain value. 

The angular distributions of $\tilde{G}$ as a function of $\theta$ in Figure~\ref{anisotropy-uniform}(b) provide another signature of collectivity. Collectivity causes an apparent ``rotation'' of the elastic anisotropy, or in other words, collectivity changes the value of $\theta$ at which the maximum of $\tilde{G}$ is attained. Again, taking $N_\text{lobe}=2$ as an example. For a single ``peanut'' grain, the greatest value occurs at approximately $\theta = 108^\circ$ (and $288^\circ$); for a collection of ``peanut'' grains (e.g., $N_\text{grain}$ = 5 and 39), the greatest value shifts to be approximately $\theta = 96^\circ$ (and $276^\circ$). We can make a similar observation for $N_\text{lobe}=4$, where the values of $\theta$ at which $\tilde{G}$ is the greatest, shift consistently by becoming $30^\circ$ smaller when going from a single grain to multiple grains. It is also expected that when $N_\text{lobe}$ is either small enough (equal to 0) or large enough (equal to 42), $\tilde{G}$ shows a nearly circular shape, indicating the cessation of elastic anisotropy.

In view of these observations, we draw the following two points for systems with a uniform $\theta$. First, individually, a single grain can show a considerable elastic softening and anisotropy when its inner void deviates from being a circle in an area-conserved manner. Second, collectively, multiple grains together enhance elastic softening while attenuating and ``rotating'',  relative to its single-grain counterpart, elastic anisotropy, both of which suggest a key role played by multi-body contact mechanics.

In all, elastic softening, stemming from simply perturbing only the geometry of a grain's inner void, serves as the dominant factor for shearing more easily a dense and ordered packing of grains under a constant solid fraction -- a seemingly very constrained and hard-to-shear setting. This finding lays the ground for studying the effect of $A$, which we elaborate on in the next Section.

\subsection{Randomly oriented grains: Amplitude-controlled shear softening}\label{result_random}
We next examine randomly oriented grain packings, which provide a more disordered and fundamentally more interesting setting, akin to the geometric disorder in granular media. To this end, we focus on a 39-grain system and study the effect of $A$ on the elastic softening of the system. We specifically probe five different values of A ($0.02R$, $0.08R$, $0.16R$, $0.22R$, and $0.3R$), recalling that while we change $A$, the area of the void remains constant, rendering all systems directly comparable. For each $A$, we consider 16 different values for $N_\text{lobe}$ ranging from 2 to 42, and average $\tilde{G}$ over five independent random orientation configurations.

Intuitively, when $A$ is small, the geometric perturbation added to the circular void is small, and the deviation of the mechanical response compared to the reference design should also be small as a consequence. This is confirmed by looking at data corresponding to $A=0.02$ (data in yellow) in Figure~\ref {anisotropy-G-correlation}(a), where $\langle G \rangle_{\boldsymbol{\theta}}$ remains largely a constant equaling to one. However, as  $A$ increases, we observe a rapid decrease $\langle G \rangle_{\boldsymbol{\theta}}$ with the increase of $N_\text{lobe}$, signifying the effect of elastic softening. For all the rest four cases (in terms of $A$), the variation of $\langle G \rangle_{\boldsymbol{\theta}}$ against $N_\text{lobe}$ shows a similar trend: a rapid decrease when $N_\text{lobe} < 10$ and a slow relaxation to an A-dependent constant value as $N_\text{lobe}$ further increases. Notably, for the greatest $A$ probed, the converged $\langle G \rangle_{\vec{\theta}}$ shows almost a $90\%$ reduction compared to that of the reference design, rendering the packing much easier to shear compared to a packing of ``donut'' grains ($A = 0$, $N_\text{lobe}=0$). It is worth emphasizing again that this reduction is achieved without changing the void area, indicating that shape, rather than porosity, is responsible for the softening. The dependence on $A$ can be understood from the local geometry of the void. At large $N_\mathrm{lobe}$, the perturbed void develops many narrow, angularly distributed, ``crack-like'' leaves (see Figure~\ref{anisotropy-G-correlation}(a) for example schematics of grains). These features concentrate strain near their tips and make the grain more compliant under shear. Increasing $A$ amplifies this effect, producing stronger grain-scale softening and, consequently, a softer packing response. Although not considered in this work, it would certainly be interesting to study whether there is a limiting value of $\langle G \rangle_{\boldsymbol{\theta}}$ as $A$ keeps increasing.

\subsection{Contact-force anisotropy}
\begin{figure}
\centering
\includegraphics[width=\linewidth]{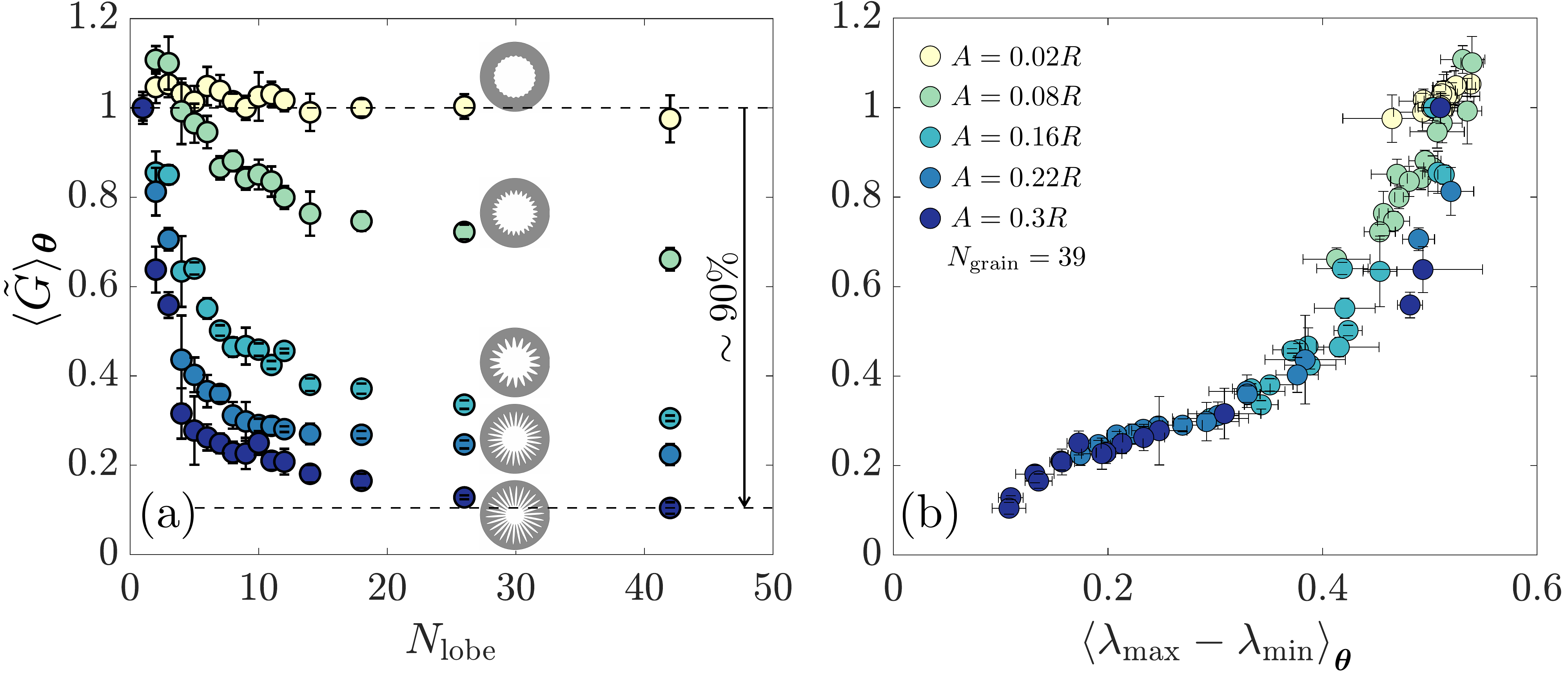}
\caption{(a) The variation of $\langle \tilde{G} \rangle_{\boldsymbol{\theta}}$, obtained by averaging over five configurations parametrized by $\boldsymbol{\theta}$, as a function of $N_\text{lobe}$. Error bars suggest the degree of variation of $\tilde{G}$ over the five configurations. Results are shown for five values of the amplitude parameter: $A = 0.02R$ (light yellow), $A = 0.08R$ (light green), $A = 0.16R$ (cyan), $A = 0.22R$ (blue), and $A = 0.3R$ (dark blue). A schematic of each grain with a given $A$ is also shown at $N_\text{lobe} = 28$. (b)The same $\langle \tilde{G} \rangle_{\boldsymbol{\theta}}$ data as in (a), but now plotted as a function of the packing's normalized contact force anisotropy $\lambda_\text{max}-\lambda_\text{min}$, averaged over the five configurations. Horizontal error bars indicate the corresponding variation of $\lambda_\text{max}-\lambda_\text{min}$ with respect to its mean value. N$_\text{grain}$ = 39 for all cases.}
\label{anisotropy-G-correlation}
\end{figure}

Finally, to connect the macroscopic shear modulus to the internal force network, we compute the normalized contact-force anisotropy. Specifically, similar to ~\cite{azema2014internal,li2020identifying}, we define the latter by the difference between the maximum ($\lambda_\text{max}$) and minimum ($\lambda_\text{min}$) eigenvalues of the following second-order symmetric tensor:
\begin{align}
\chi = \sum_{i=1}^{N_\text{c}}\frac{f_i}{\sum_{j=1}^{N_\text{c}}f_j} \mathbf{n}_i \otimes \mathbf{n}_i,
\end{align}
where $N_c$ is the total number of contact in the system, $f_i$ is the contact force magnitude of the $i$th contact, and $\mathbf{n}_i$ is the corresponding contact force direction. Each contact is weighted by the amount of contact force it carries, normalized by the total amount of contact force carried by all contacts in the system. This way, the eigenvalues of $\chi$ are bounded from above by 1. Physically, the two eigenvectors of $\chi$ indicate, in an average sense, the two most popular yet orthogonal contact force directions in a system. So, a greater value of $\lambda_\text{max}-\lambda_\text{min}$ indicates a more preferred contact force direction in the system, which, in the limit of quasi-statics, implies the formation of percolating force chains in resisting the applied shear and subsequently a non-zero apparent shear modulus. Figure~\ref{anisotropy-G-correlation}(b) collects all $\langle G \rangle_{\boldsymbol{\theta}}$ data for the 39-grain system in Figure~\ref{anisotropy-G-correlation}(a) and plots them with $\langle\lambda_\text{max}-\lambda_\text{min}\rangle_{\boldsymbol{\theta}}$. All data collapse fairly well together, confirming the positive correlation between macro-scale ($\langle G \rangle_{\boldsymbol{\theta}}$) and particle-scale ($\langle\lambda_\text{max}-\lambda_\text{min}\rangle_{\boldsymbol{\theta}}$) observations.

In all, here we show that the elastic softening effect found from a packing with uniformly oriented grains carries over to a packing with randomly oriented grains. More importantly, depending on the magnitude of $A$ and $N_\text{lobe}$, an ordered and dense packing of grains can be much easier to shear compared to the reference case, and again, without involving particle-scale finite deformation. We also find our conclusion robust against variations in contact friction (see Figure~\ref{add-friction} where we vary $\mu_\text{p}$ to be either 0.5 or 0.9) and system size (see Figure~\ref{add-systemsize} where we further consider a 137-grain system).

\section{Summary}\label{discussion}
In this work, through detailed finite-element simulations and analysis, we show that shearing even an ordered and dense granular packing can be made substantially easier by \textit{only} changing the internal geometry of grains, without involving particle-scale auxeticity or finite deformation. Starting from annular grains with circular voids, we introduced area-preserving perturbations to the void's geometry. Fundamentally, this geometry-induced anisotropy, paired with the elasticity of the constituent material, leads to an apparent elastic softening and anisotropy of a single grain. When such grains are assembled into dense packings, multi-body contact interactions transmit this grain-scale effect to the packing scale, leading to a strong reduction in the apparent shear modulus. Our work highlights how grain-scale geometry, when mediated by multi-body contact mechanics, can modulate system-scale elasticity, providing a simple route for designing granular metamaterials with targeted elastic properties through geometrical design alone. 

Several questions remain open. First, the present work focuses on small applied shear strains and linear elastic grain response. It would be particularly interesting to study how our system would behave under large shear strains, as in \cite{poincloux2024rigidity}. At such larger strains, the repeated breaking and re-forming of contacts observed in our simulations may continue to suppress the formation of stable force chains, or the packing may instead transition toward collective stiffening. Furthermore, we applied a modest pressure, and each grain deforms linearly in response. What if we apply a large enough pressure such that most grains in the system develop finite deformation before shear commences? Exploring these possibilities will require simulations with geometric and material nonlinearities as well as physical experiments. Experiments would also make it possible to test the role of basal friction and other three-dimensional effects that are not captured in the present two-dimensional model. 

\section*{CRediT authorship contribution statement}
\textbf{Liuchi Li}: Conceptualization, Software, Formal analysis, Visualization, Validation, Investigation, Methodology, Funding acquisition, Project administration, Writing - Original Draft, Writing - Review \& Editing. \textbf{Konstantinos Karapiperis}: Conceptualization, Investigation, Methodology, Funding acquisition, Project administration, Writing - Review \& Editing.

\section*{Declaration of competing interest}
The authors declare that they have no known competing financial interests or personal relationships that could have appeared to
influence the work reported in this paper.

\section*{Data availability}
Data will be made available upon reasonable request.

\section*{Acknowledgments}
L. L. and K. K. gratefully acknowledge the financial support through the  Scientific Exchange Grant from the Swiss National Science Foundation as well as the computing resources provided by EPFL's SCITAS cluster, both of which contributed to the early-stage development of this work. L. L. is pleased to acknowledge that the work reported on in this paper was substantially performed using the Princeton Research Computing resources at Princeton University. Princeton Research Computing is a consortium of groups, including the Princeton Institute for Computational Science and Engineering (PICSciE) and Research Computing at Princeton University. L. L. also gratefully acknowledges Princeton University and the Department of Civil and Environmental Engineering for financial support provided through the start-up funding.

\appendix
\section{}

\begin{figure}[h]
\centering
\includegraphics[width=\linewidth]{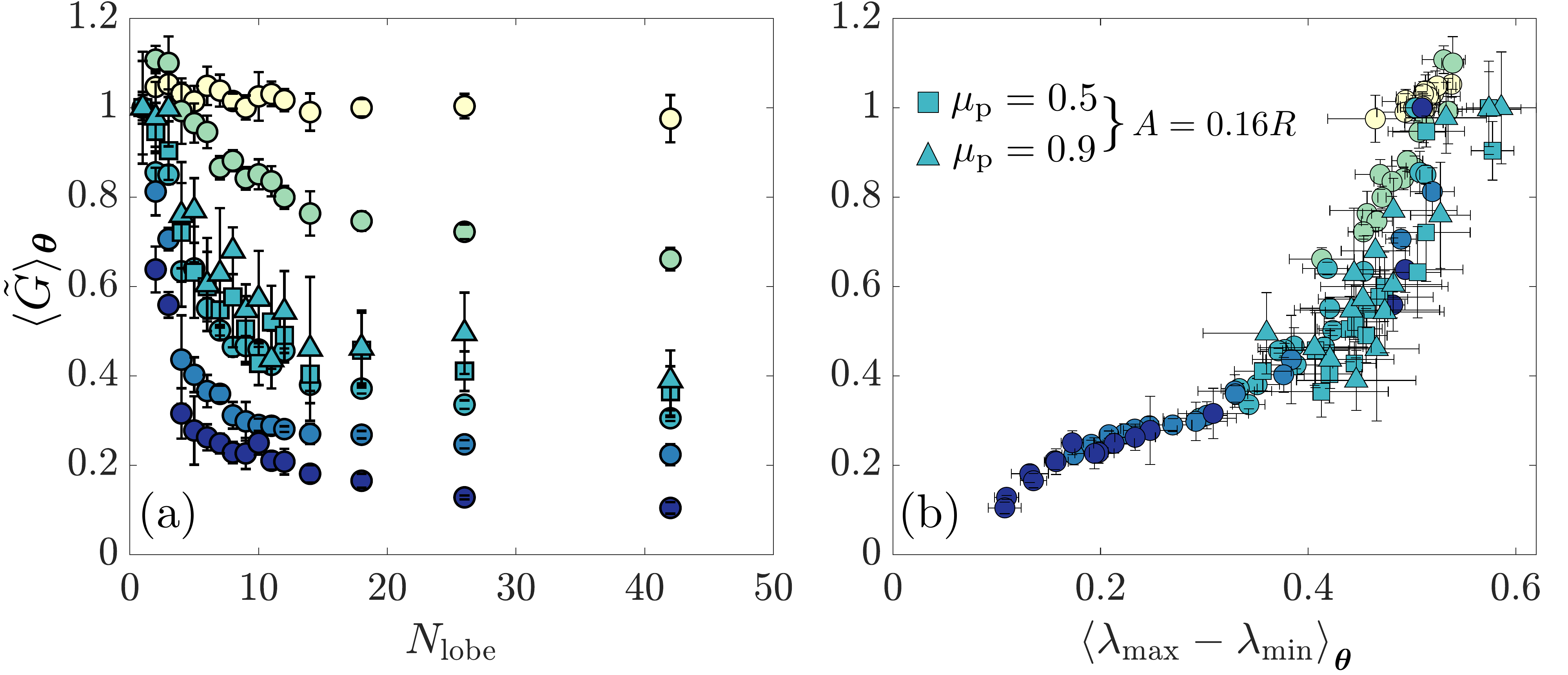}
\caption{A replication of Figure.~\ref{anisotropy-G-correlation}, but now with the addition of results from $\mu_\text{p} = 0.5$ ( squares in cyan) and $\mu_\text{p} = 0.9$ (upper triangles in cyan).}
\label{add-friction}
\end{figure}  

\begin{figure}[h]
\centering
\includegraphics[width=\linewidth]{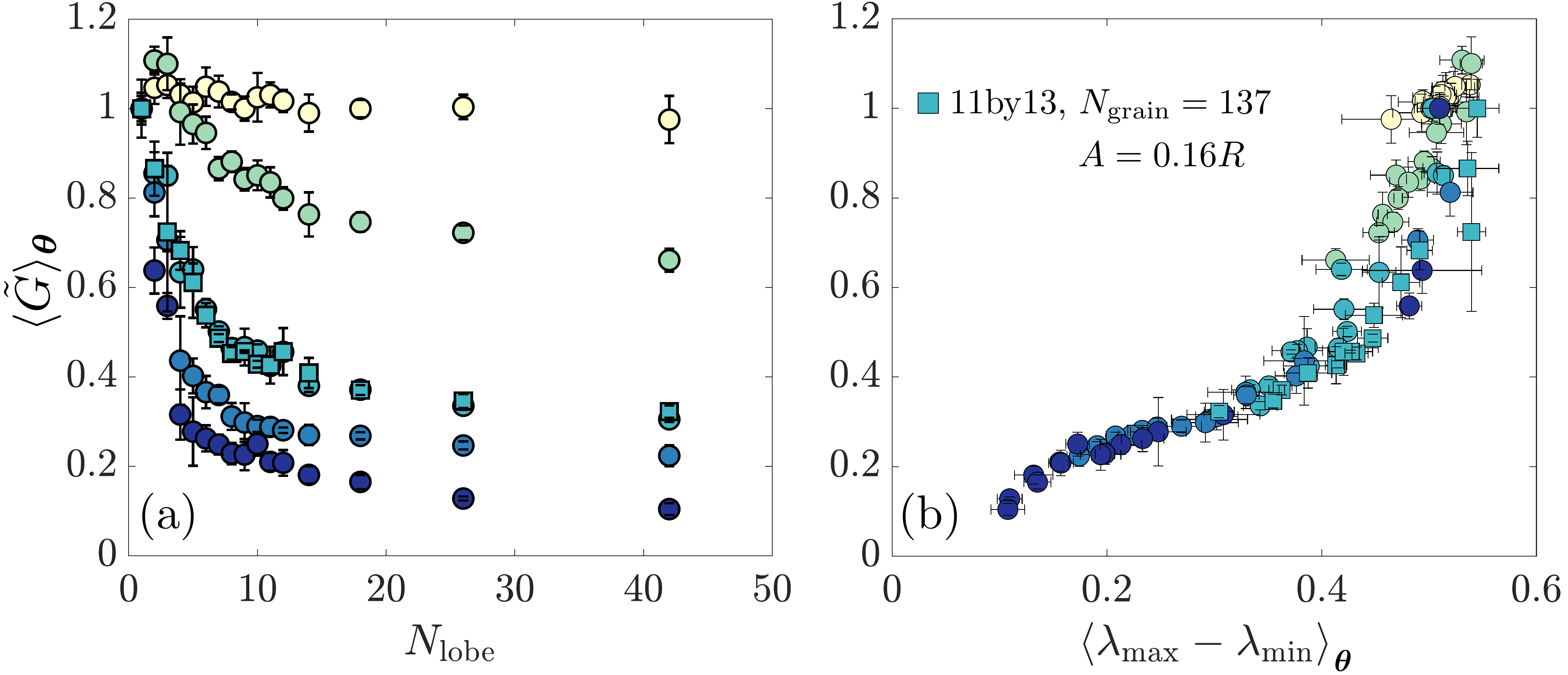}
\caption{A replication of Figure.~\ref{anisotropy-G-correlation}, but now with the addition of results from $N_\text{grain} = 137$ (cyan square).}
\label{add-systemsize}
\end{figure}  

\bibliographystyle{unsrt}
\bibliography{ref}

\end{document}